\documentclass{article}

\usepackage[final]{nips_2017}

\usepackage[utf8]{inputenc} 
\usepackage[T1]{fontenc}    
\usepackage{hyperref}       
\usepackage{url}            
\usepackage{booktabs}       
\usepackage{amsfonts}       
\usepackage{nicefrac}       
\usepackage{microtype}      
\usepackage{graphicx}		

\usepackage[utf8]{inputenc}
 
\usepackage{listings}
\usepackage{color}
 
\definecolor{codegreen}{rgb}{0,0.6,0}
\definecolor{codegray}{rgb}{0.5,0.5,0.5}
\definecolor{codepurple}{rgb}{0.58,0,0.82}
\definecolor{backcolour}{rgb}{1,1,1}
\definecolor{orange}{rgb}{1,0.50,0}
 
\lstdefinestyle{mystyle}{
    backgroundcolor=\color{backcolour},   
    commentstyle=\color{codegray},
    keywordstyle=\color{orange},
    numberstyle=\tiny\color{codegray},
    stringstyle=\color{codegreen},
    keywordstyle=[2]\color{codepurple},
    basicstyle=\footnotesize,
    breakatwhitespace=false,         
    breaklines=false,                 
    captionpos=b,                    
    keepspaces=true,                 
    numbers=left,                    
    numbersep=5pt,                  
    showspaces=false,                
    showstringspaces=false,
    showtabs=false,                  
    tabsize=2
}
 
\title{Machine Learning Techniques for Brand-Influencer Matchmaking on the Instagram Social Network}

\author{
  Taylor Sweet \\
  Department of Mechanical Engineering\\
  University of British Columbia\\
  \texttt{taylor.sweet@alumni.ubc.ca} \\
   \And
   Austin Rothwell \\
   Department of Computer Science \\
   University of British Columbia \\
   \texttt{acroth@cs.ubc.ca} \\
   \And
   Xuan Luo \\
   Department of Electrical and Computer Engineering \\
   University of British Columbia \\
   \texttt{xuanluo2@ece.ubc.ca} \\
}

\begin{document}

\maketitle

\begin{abstract}
  The social media revolution has changed the way that brands interact with consumers. Instead of spending their advertising budget on interstate billboards, more and more companies are choosing to partner with so-called Internet "influencers" --- individuals who have gained a loyal following on online platforms for the high quality of the content they post. Unfortunately, it's not always easy for small brands to find the right influencer: someone who aligns with their corporate image and has not yet grown in popularity to the point of unaffordability. In this paper we sought to develop a system for brand-influencer matchmaking, harnessing the power and flexibility of modern machine learning techniques. The result is an algorithm that can predict the most fruitful brand-influencer partnerships based on the similarity of the content they post. 
\end{abstract}

\section{Introduction}
The social media age has led to the creation of a near-infinite and readily-accessible source of data regarding users' habits, interests, and brand-inclinations. To date, this trove of information has been used to make personal product recommendations, serve targeted ads, and generally gain insight into the mind of the consumer. However, despite the meteoric rise of certain Internet celebrities, earning them coveted roles as ambassadors for some of the world's biggest brands, there still remains a large community of indie creators with untapped potential for brand partnerships.

The rise of the social network has caused a paradigm shift in celebrity culture and has redefined what it means to be ``famous''. In the early days of social media the users that had the biggest followings tended to be those who had achieved fame in one or more traditional capacities, such as: business, politics, or entertainment, to name a few. Nowadays, however, there exists a sizable number of content creators, spanning platforms such as Facebook, Youtube, and Instagram, that have achieved Internet fame, with follower/subscriber numbers in the millions.

With the ease of Internet marketing we have seen an increasing number of brands forming partnerships with these so-called ``Internet celebrities'', from Kylie Jenner and Adidas to Felix Kjellberg (Youtube's PewDiePie) and Disney, there is no shortage of big-time deals that have been facilitated by the rise of the social network. However, what is often overlooked is the number of creators with small to medium sized audiences (10-100K) who have achieved notoriety in their own regard. While their individual audience sizes may be smaller, the up-and-coming food, fashion, and lifestyle bloggers, amongst countless others, in aggregate, have the potential to influence the purchasing decisions of a large proportion of the population. Moreover, a partnership with these indie influencers would be much more affordable for a smaller brand than the likes of Kim Kardashian, who charges in the neighbourhood of \$500,000 for an ad spot [1].

If the wealth of data that exists on social media were to be harnessed, it could, with a little ingenuity, be used to facilitate brand-influencer matchmaking. Not only would this help companies find content creators that align with their brand image, it would also provide an opportunity for the small-time creators to monetize their posts, further encouraging the creation of high-quality future content.

While many researchers have published papers on social media, only a select few have addressed the role that influencers play on the platforms, and those that have either didn't take a machine learning approach or neglected to propose a method for matching brands with influencers and vice-versa. Thus, this paper seeks to fill this knowledge gap by presenting new research into the application of machine learning techniques for brand-influencer matchmaking on social media.

The specific contributions of this work are:
\begin{enumerate}
\item A first-of-its-kind tool for performing content analysis on a user's Instagram profile.
\item A novel algorithm for matching brands with social media influencers.
\end{enumerate}

\section{Related Work}
A considerable number of researchers have addressed the concept of mining and manipulating social media user data. With emphasis placed on a variety of platforms, such as Twitter, Pinterest and Instagram, researchers have sought answers to questions such as: ``What types of users exist on the platform?'', ``What characteristics are common amongst influencers?'', and ``What factors drive interactions with a particular piece of content?''.

In an effort to address the ``user type'' query, Hu et al. [2] crawled data from the Instagram social network, collecting the 20 most recent photos from each of 50 randomly selected users. They then used computer vision techniques such as Scale Invariant Feature Transform (SIFT) to detect and extract local discriminative features from the images. Finally, K-means clustering was used to group the images into distinct categories and the users into particular classes. The authors conclude that there are 15 general types of photos posted on Instagram, which they manually narrow down to 8 categories (friends, food, gadget, text, pet, activity, self-portrait, and fashion), and 5 general user types (no labels were given to the user type clusters, though some, such as ``selfie lover'', were inferred).

In their 2011 paper, ``A Machine Learning Approach to Twitter User Classification'', Pennacchiotti and Popescu [3] also sought to use social media post characteristics to classify users. In this study, however, the platform of choice was Twitter. By leveraging observable information such as the user behavior, network structure, and the linguistic content of the user's Twitter feed, they were able to automatically infer the values of user attributes such as political orientation, ethnicity, and brand affinity. This was achieved through the use of the unsupervised learning algorithm Gradient Boosted Decision Trees (Friedman 2001), which consists of an ensemble of decision trees fitted in a forward stepwise manner to current residuals. Using this algorithm the authors were able to decide whether a random user was a potential follower of the Starbucks coffee chain with 76.3\% accuracy, with results indicating that profile and linguistic information were the most helpful features.
				
Tangential to the topic of user type classification is the concept of user interest classification, which is a research topic addressed by Xie et al. [4] in their paper: ``Mining User Interests from Personal Photos''. In this work 180,000 pictures are mined from the Flickr photo-sharing service, comprising the profiles of 227 distinct users. For the purpose of user interest classification a User Image Latent Space Model is used to jointly model user interests and image contents, where user interests are modeled as latent factors and each user is assumed to have a distribution over them. Image contents are modeled by a four-level hierarchical structure where the layers correspond to themes, semantic regions, visual words and pixels, respectively. Given the image contents a user's interests could be discerned by using a variational inference method. In comparison to the baseline methods of K-means clustering and linear discriminant analysis (LDA) Xie et al.'s algorithm was found to have better classification precision (0.6907) versus K-means (0.5563) and LDA (0.5940).

Among the first to investigate the driving force behind social media brand interactions were Singh et al. [5]. In this work an attempt was made to discern what media content, in a predictable manner, can generate high interaction rates amongst a brand's target audience, captured through ``repinnings'' on Pinterest. In this study, user and brand image data is represented using a bag-of-visual-words model (BoVW), which creates an image dictionary by taking as input a set of image-based features and using them to generate a set of visual words through clustering. A similarity measure can then be computed between a brand's pin and all the pins of each user. The prototype system presented in this paper is able to predict who will interact with a brand image on Pinterest by ranking users based on the similarity measure and inferring their propensity to interact with that particular pin.  The system was tested on a large-scale dataset of more than 1 million images and produced ``promising results'' --- a vague but intriguing claim by the authors.	
			
One sure-fire way to drive content interactions is for brands to partner with social media influencers. Unfortunately, however, it's not always easy for businesses to find the right person to market their products --- someone who upholds the values of the company, has an appreciable audience	 size, and posts high-quality content. In an effort to address this issue, Booth and Matic [6] propose an algorithm that provides a score (`influencer index') for how valuable a blogger could be to a particular company. The procedure is to choose some users and manually score them based on certain criteria (e.g., posts per month, views per month, etc.) and then use these results to determine their `influencer index'. The issue with this approach is that all of the scoring is done manually and no machine learning techniques are implemented to facilitate the brand-influencer matchmaking. 

In a separate study focused on the presence of influencers on social media Lahuerta-Otero and Cordero-Guti\'errez [7] attempted to determine what characteristics define an influencer, with particular focus placed on Twitter. For this purpose a unique data mining tool was used that combines graph theory and social influence theory. An analysis of 3853 users posting about two Japanese automotive manufacturers, Toyota and Nissan, revealed the characteristics influencers have on this social network. The findings suggested that influencers use more hashtags and mentions on average when they tweet, with a smaller word count and fewer embedded links in their posts. Additionally, influencers tend to follow a large number of people themselves and don't shy away from expressing their feelings, either positive or negative, when tweeting.

\section{Our Approach}

In this paper brand-influencer matchmaking is facilitated by first conducting an initial analysis of the content on a user's Instagram feed, as well as that of the target brand. Next, a machine learning algorithm is implemented to determine the similarity between profiles and make predictions regarding the most potentially fruitful partnerships. Finally, the results are visualized and the findings discussed in detail.

\subsection{User Profile Content Analysis}
\subsubsection{Data Crawling}
In order to download information from a user's Instagram profile we made use of an open-source tool developed in Python by Richard Arcega [8]. This tool facilitates the download of all media content uploaded to a specified user's Instagram account, as well as the associated captions and hashtags. Using this approach, we downloaded the profiles of 20 unique users, spanning 5 content themes: dogs, cats, mountains, cars, and pizza. We also downloaded the profiles of a variety of brands which each serve markets that fall into one of these categories. 

\subsubsection{Object Recognition}

Armed with the media library of a variety of users, the next task was to analyze the content in order to extract common themes. For this purpose, a TensorFlow image classification algorithm was used (Inception-v3), which was trained on the ImageNet database [9]. The output of this algorithm is a list of the 5 most likely tags for a given image as well as an associated confidence score for each. These tags were then appended on an image-by-image basis to the metadata files produced by the Instagram scraper.

\subsection{Matching Brands to Influencers}

\subsubsection{Data Synthesis and Pre-processing}

In order to see the ``bigger picture'' with regards to a user's posting habits and the overarching themes of their content, it was necessary to first synthesize the image classification results from each individual post. This was accomplished by assembling the three most likely tags for each image into a long `profile content string' for each user. The profile content strings were then compiled into a global matrix. 

Since most machine learning algorithms expect numerical features as an input, the global matrix of user profile content strings had to be vectorized. For this purpose, a "Bag of Words" strategy was used. This approach involves three steps: tokenizing words and giving an integer identifier to each possible token, counting the occurrence of tokens in each user string, and normalizing and weighting with diminishing importance the tokens that appear in the majority of user strings. The result of this vectorization is an $m$ by $n$ matrix, where $m$ is the number of users and $n$ is the number of unique image tags. The entry at any given position in the matrix reflects the frequency of occurrence of the image tag (column) in that user's profile (row). 

\subsubsection{Making Predictions}

The final and most important step in the matchmaking workflow is to predict which influencers have profiles that are most closely aligned to that of a particular target brand. In order to facilitate this matchmaking objective the user profile content string associated with the target brand was incorporated into the global matrix of feature vectors. Next, a k-Nearest Neighbours (k-NN) model was trained using the combined matrix of influencer and target brand profile data as input.

With a trained k-NN model, predictions could be made based on the proximity of influencer profiles to the target brand profile, using Euclidean distance as the similarity metric. In this way, it was possible to determine which influencers would be a good fit for a particular brand based on the likeness of the content they post.

\section{Experimental Results}

In order to verify our approach and validate the model we tested its performance on a variety of sample brands. For this purpose we selected 1-2 companies in each of three different markets: dog food, outdoor goods,  and pizza delivery. The selected brands were: \textit{Mountain Dog Food}, \textit{Mountain Hardwear}, \textit{Giordano's Pizza}, and \textit{Domino's Pizza}. 

Our metric for performance evaluation was the perceived sensibility of the cluster of profiles to which the target brand was assigned. For instance, we would expect that the dog food brand be assigned to the cluster of influencers that post about dogs, rather than about mountains or pizza. Using Multi-Dimensional Scaling (MDS) as a technique to visualize the high-dimensional data, our algorithm produced the following results.  

\begin{figure}[h]
  \centering
  \includegraphics[keepaspectratio,height=9cm, width=9cm]{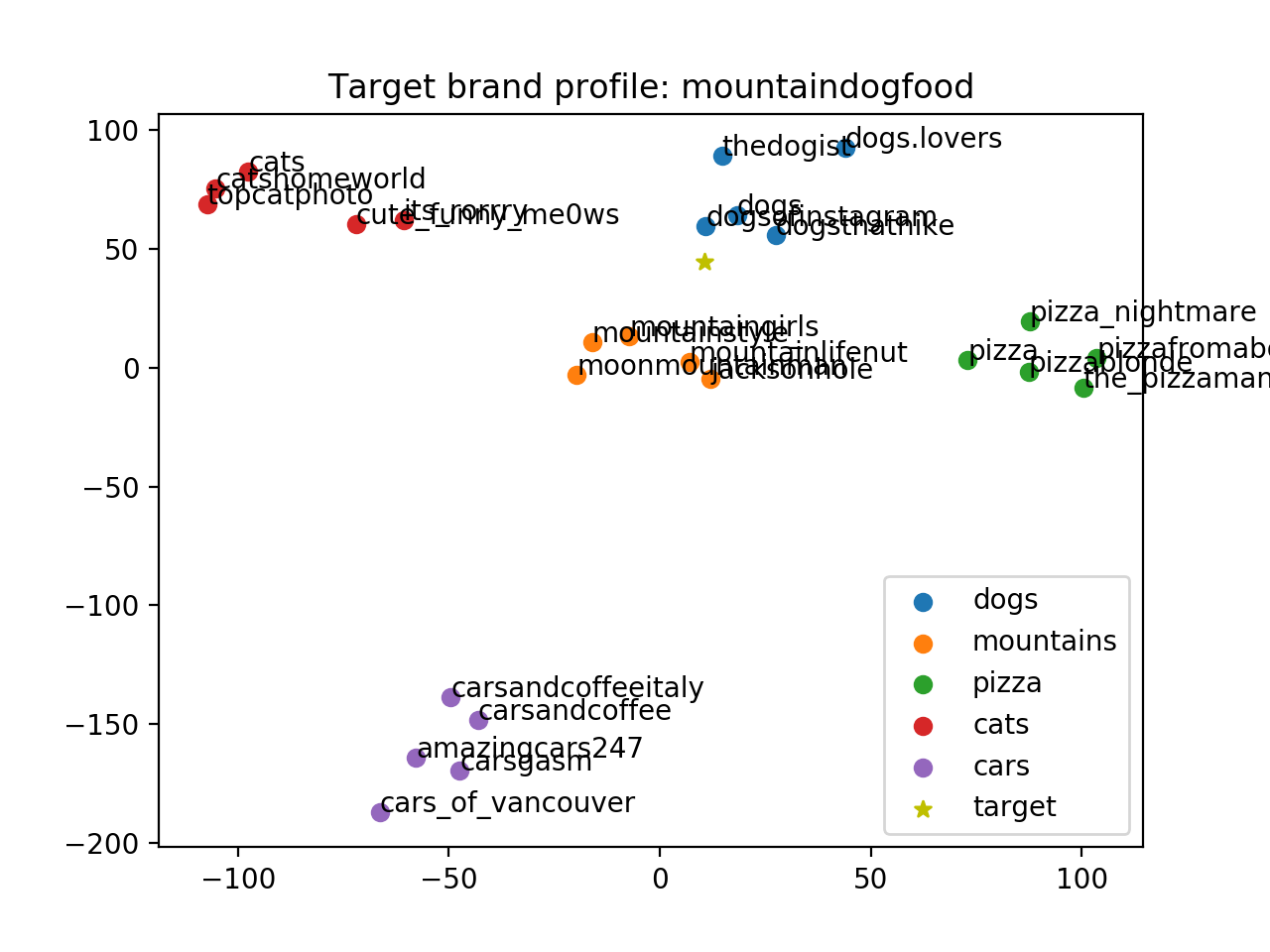}
  \caption{MDS plot with `\textit{mountaindogfood}' as target brand profile.}
\end{figure}

\begin{figure}[h]
  \centering
  \includegraphics[keepaspectratio,height=9cm, width=9cm]{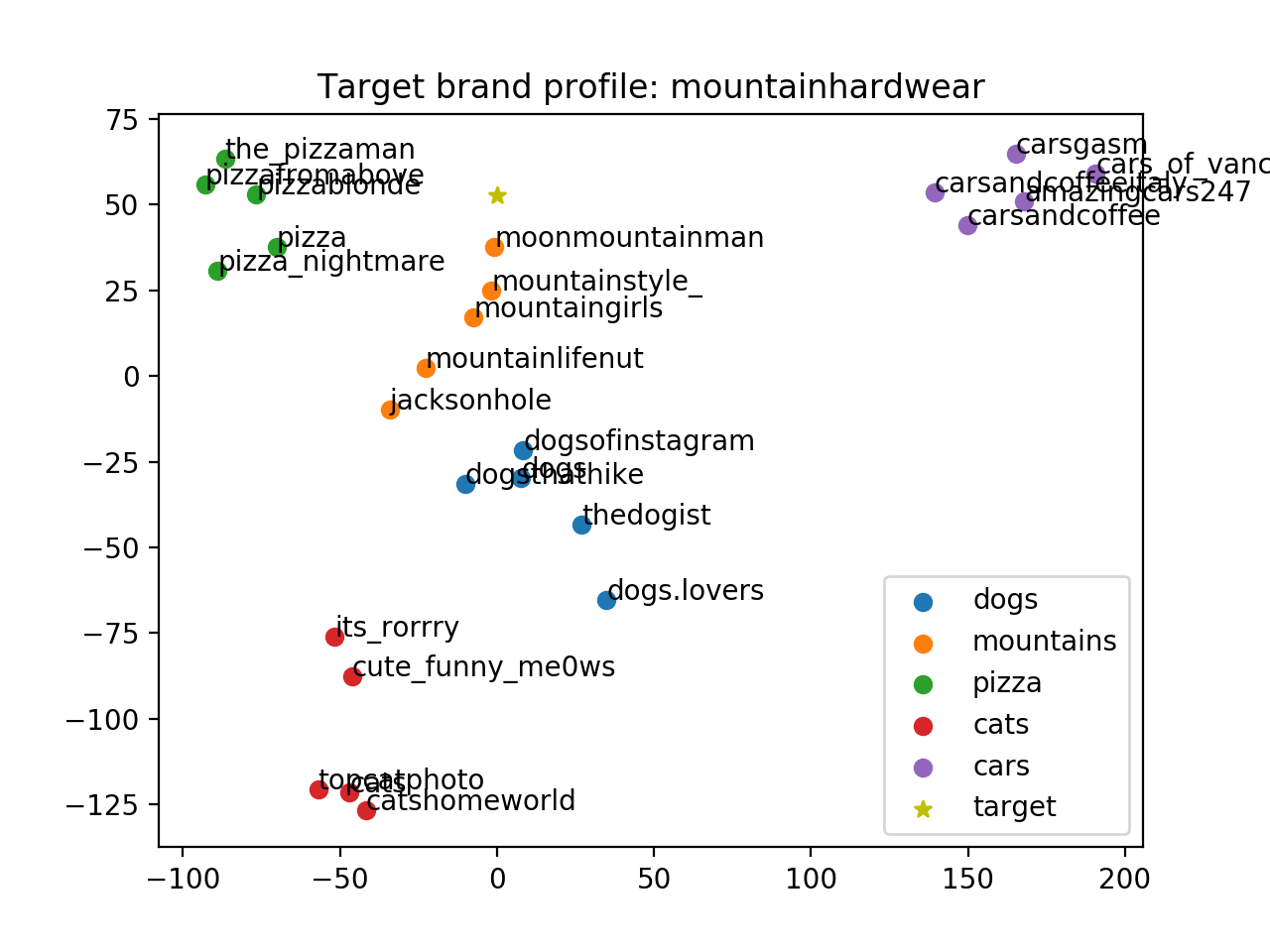}
  \caption{MDS plot with `\textit{mountainhardwear}' as target brand profile.}
\end{figure}

As we can see in Figure 1, the target profile, `\textit{mountaindogfood}', lies closest to the cluster of profiles that post dog-related content. Similarly, in Figure 2, the target profile `\textit{mountainhardwear}', lies closest to the other mountain-oriented profiles.

\newpage

These results are well in line with expectation, based on the content of the influencer profiles and that of the target brands. The more informative result, however, is not the grouping of the target profile, but rather its relative position within the group. Based on the distribution of profiles within the mountain cluster (shown in orange), `\textit{moonmountainman}' is the user that posts content that is most similar to \textit{Mountain Hardwear}. Thus, we would recommend that \textit{Mountain Hardwear} partner with this influencer above all others in order to extend their marketing reach while ensuring that they remain on target demographically.

Comparing two companies that operate within the same market segment, \textit{Giordano's Pizza} and \textit{Domino's Pizza}, we see that our algorithm would predict a different ordering for the most appropriate influencers for these two brands (see Figure 3). For \textit{Giordano's Pizza}, the nearest neighbours, in descending order, are: `\textit{pizza}', `\textit{pizzablonde}', `\textit{pizza\textunderscore nightmare}', `\textit{the\textunderscore pizzaman}', and `\textit{pizzafromabove}'. On the other hand, those for \textit{Domino's Pizza} are: `\textit{pizzafromabove}', `\textit{pizzablonde}', `\textit{the\textunderscore pizzaman}', `\textit{pizza}', and `\textit{pizza\textunderscore nightmare}'. 

\begin{figure}[h]
  \centering
  \includegraphics[keepaspectratio,height=5cm]{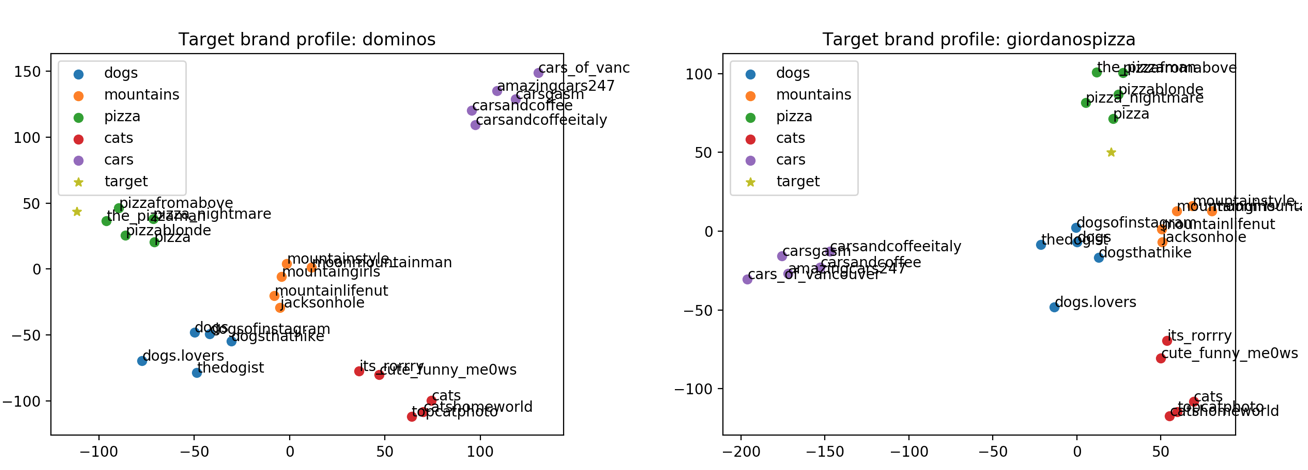}
  \caption{MDS plot with target brand proflies: `\textit{dominos}' (left) and `\textit{giordanospizza}'  (right).}
\end{figure}

\section{Discussion of Findings}

In this study we sought to match brands with influencers on Instagram using machine learning techniques. Specifically, we first scraped Instagram user data and analyzed the content of user images with a TensorFlow neural network. Then, the processed user data was vectorized using a ``Bag of Words'' approach. Finally, k-NN was used to predict which Instagram users had profiles that were most similar to those of particular target brands.

Our algorithm is able to provide a target brand with a list of Instagram users who may be suitable advertising partners. This is a useful tool for brands seeking exposure on Instagram, considering the millions of users on the platform and that only a small proportion of users post photos relevant to a given brand. Further, these users can be difficult to find based on traditional search information, such as usernames, hashtags, or captions. Importantly, our matching algorithm does not use such metadata to find influencers. Rather, content analysis is performed on posted images to acquire in-depth information on the posting behaviour of each user, which is subsequently used to suggest brand-influencer partnerships.

Though our algorithm succeeds at determining brand-influencer similarity in our test dataset, there are limitations to our approach. The success of our algorithm was demonstrated in a contrived user environment that consisted of five broad categories of users. As such, the performance of the algorithm has not been validated on the broader extent of the Instagram user-base, which contains more users, which fall into less distinct categories. In addition, our predictions rely on the validity of the content analysis that is performed on each image. While the TensorFlow Inception-v3 neural network is state-of-the-art at the time of writing, future improvements to image content analysis should be integrated to ensure best performance.

There are multiple other improvements that could be made to the algorithm described herein in order to improve the efficacy of brand-influencer matchmaking. For instance, metadata could be extracted from the influencer profiles and used to create a more complete depiction of the individual. Examples of such metadata include: number of followers, average number of likes per post, and posting frequency. Though posted images, for the most part, define the theme of an Instagram user, such supplementary information would allow brands to narrow the terms of their search in order to target influencers with, for example, a specific number of followers or threshold level of content interaction. 

Another potential improvement involves basing brand-influencer matchmaking on custom parameters, rather than the Instagram profile of the brand. For example, a brand like \textit{Domino's Pizza} may want to be matched with influencers who post photos of many types of food, instead of exclusively pizza. To enable this, \textit{Domino's} could identify general image themes of interest (e.g., food, drink, sports), which would then be narrowed down to particular image content descriptors, vectorized using ``Bag of Words'', and used to find the appropriate influencers. This would allow a brand to specify the type of user that is desired in the case where the content they post differs from that of the brand's own Instagram profile.

\section{Conclusion}

In this paper we have presented a novel approach to brand-influencer matchmaking based on well-known machine learning techniques and using an image-based analysis of user profile content. The results indicated that our algorithm, when presented with a variety of potential influencer profiles, is able to identify those profiles that are most closely aligned with a particular target brand. Improvements could be made to this approach by incorporating more forms of user data, using a higher-fidelity image analysis engine, and allowing for manual input of what a brand is looking for in an influencer.  

\newpage

\section*{References}

\small

[1] S. Mohammed, ``The Kardashians make HOW much for a sponsored Instagram post?'', Glamourmagazine.co.uk, 2018. [Online]. Available: http://www.glamourmagazine.co.uk/article/cost-of-kardashians-instagram-posts. [Accessed: 27- Mar- 2018].

[2] Y. Hu, L. Manikonda and S. Kambhampati, ``What We Instagram: A First Analysis of Instagram Photo Content and User Types'', in Eighth International AAAI Conference on Weblogs and Social Media, 2014, pp. 595-598.

[3] M. Pennacchiotti and A. Popescu, ``A Machine Learning Approach to Twitter User Classification'', in Fifth International AAAI Conference on Weblogs and Social Media, 2011.

[4] P. Xie, Y. Pei, Y. Xie and E. Xing, ``Mining User Interests from Personal Photos'', in Twenty-Ninth AAAI Conference on Artificial Intelligence, 2015.

[5] S. Singh, Y. Wang and L. Ding, ``Who is repinning? Predicting a brand's user interactions using social media retrieval'', in MDMKDD, Chicago, Illinois, USA, 2013.

[6] N. Booth and J. Matic, ``Mapping and leveraging influencers in social media to shape corporate brand perceptions'', Corporate Communications: An International Journal, vol. 16, no. 3, pp. 184-191, 2011.

[7] E. Lahuerta-Otero and R. Cordero-Guti\'errez, ``Looking for the perfect tweet. The use of data mining techniques to find influencers on twitter'', Computers in Human Behavior, vol. 64, pp. 575-583, 2016.

[8] R. Arcega, ``Instagram Scraper'', GitHub, 2017. [Online]. Available: https://github.com/rarcega/instagram-scraper. [Accessed: 11- Apr- 2018].

[9] M. Abadi, A. Agarwal, P. Barham, et al.
``TensorFlow: Large-scale machine learning on heterogeneous systems'',
2015. Software available from tensorflow.org.

\newpage

\section{Appendix: Code}

\begin{lstlisting}[basicstyle=\tiny, language=Python, caption=Data crawling and object identification code.,morekeywords={range}]
import os
import subprocess
import matplotlib.pyplot as plt
import matplotlib.image as mpimg
import json
import re


os.environ['TF_CPP_MIN_LOG_LEVEL'] = '2'   # To silence some TensorFlow warnings

# Load Instagram user list
instagram_users = open('instagram_users3.txt').read().split('\n')

for username in instagram_users:

    if not os.path.isfile('./users/' + username + '.json'):
        n_posts = 100   # Total number of posts to scrape (images and videos)
        n_images = 50   # Number of images you want to analyze

        # Scrape Instagram profile for n_images most recent posts and collect metadata
        os.system('instagram-scraper ' + username + ' --maximum ' + str(n_posts) +
                  ' -u austinrothwell -p ------- --media-metadata --destination ./users/' + username)

        # Directory that user images are saved in
        images_directory = './users/' + username + '/'

        # Load user metadata, remove videos, remove video metadata from metadata
        metadata = json.load(open(images_directory + username + '.json'))
        os.system('rm ' + images_directory + '*.mp4')
        [metadata.remove(post) for post in metadata if post['is_video']]

        # check if the total count of the instagram users is less than 50
        count = len(metadata)  # type: int

        if n_images > count:
            n_images = count

        metadata = metadata[0:n_images]  # Trim metadata to size of n_images

        # Classify each image and show with metadata
        for image in range(0, n_images):
            image_name = metadata[image]['urls'][0].split('/')[-1:][0]
            filename = os.fsdecode(image_name)
            if filename.endswith(".jpg"):

                # Image classification
                out = subprocess.check_output('python ./models/tutorials/image/imagenet/classify_image.py --image_file '
                                              + images_directory + filename, shell=True)

                # Get image metadata
                image_class = str(out).split('\\')[0][2:] + '\n' + str(out).split('\\')[1][1:]
                n_likes = metadata[image]['edge_media_preview_like']['count']
                n_comments = metadata[image]['edge_media_to_comment']['count']
                try:
                    caption = metadata[image]['edge_media_to_caption']['edges'][0]['node']['text']
                    tags = ' '.join(metadata[image]['tags'])
                except:
                    caption = 'n/a'
                    tags = 'n/a'

                # Write image contents to user JSON file
                metadata[image]['image_contents'] = [s.split(' (')[0] for s in str(out)[2:-3].split('\\n')]
                metadata[image]['image_scores'] = [float(re.findall("\d+\.\d+", s)[0]) for s in str(out)[2:-3].split('\\n')]
                json.dump(metadata, open(images_directory + username + '.json', 'w'),
                          sort_keys=True, indent=4, separators=(',', ': '))

                 # Show image and predicted labels
                 img = mpimg.imread(images_directory + filename)
                 imgplot = plt.imshow(img)
                 plt.title(image_class)
                 plt.xlabel('likes: ' + str(n_likes) + ', comments: ' + str(n_comments) + '\n' +
                            'caption: ' + caption + '\n' + 'tags: ' + tags)
                 plt.gcf().subplots_adjust(bottom=0.18)
                 plt.show()

        # Delete all images from user directory and move .json metadata to .\users
        os.rename(images_directory + username + '.json', './users/' + username + '.json')
        os.system('rm ' + images_directory + '*.jpg')
        os.rmdir(images_directory)
\end{lstlisting}

\newpage

\begin{lstlisting}[basicstyle=\tiny, language=Python, caption=Data analysis code.]
import json
from sklearn.feature_extraction.text import CountVectorizer
from sklearn.neighbors import NearestNeighbors
import matplotlib.pyplot as plt
from sklearn.manifold import MDS

# Directory that metadata is saved in
metadata_directory = './users/'

# Load Instagram user list
instagram_users = open('instagram_users_organized.txt').read().split('\n')

user_str = ''
global_str = []

# Loop over users
for username in instagram_users:

    metadata = json.load(open(metadata_directory + username + '.json'))

    for i in range(1, len(metadata)):
        if metadata[i].get('image_contents'):
            user_str += ' ' + metadata[i]['image_contents'][0]
            user_str += ' ' + metadata[i]['image_contents'][1]
            user_str += ' ' + metadata[i]['image_contents'][2]

    global_str.append(user_str)
    user_str = ''

# Vectorize image content strings
vectorizer = CountVectorizer()
X = vectorizer.fit_transform(global_str)

# Fit K-Nearest Neighbors model
neigh = NearestNeighbors(n_neighbors=5)
neigh.fit(X)

# Specify target brand profile position in 'instagram_user' list
target = 25
X_tilde = X[target]
print('Target profile is:')
print(instagram_users[target])

# Find nearest neighbors
neighbors = neigh.kneighbors(X_tilde)[1][0]
print('\n' + 'Most closely related profiles are:')
for j in range(1, len(neighbors)):
    print(instagram_users[neighbors[j]])

# Multi-dimensional Scaling (MDS) to visualize user distance in 2D
X_lowdim = MDS(n_components=2).fit_transform(X.toarray())
ax = plt.gca()
c1 = ax.scatter(X_lowdim[0:5, 0], X_lowdim[0:5, 1], label='dogs')
c2 = ax.scatter(X_lowdim[5:10, 0], X_lowdim[5:10, 1], label='mountains')
c3 = ax.scatter(X_lowdim[10:15, 0], X_lowdim[10:15, 1], label='pizza')
c4 = ax.scatter(X_lowdim[15:20, 0], X_lowdim[15:20, 1], label='cats')
c5 = ax.scatter(X_lowdim[20:25, 0], X_lowdim[20:25, 1], label='cars')
c6 = ax.scatter(X_lowdim[target, 0], X_lowdim[target, 1], label='target', marker='*', color='y')

plt.legend(handles=[c1, c2, c3, c4, c5, c6])
plt.title('Target brand profile: ' + instagram_users[target])

# Annotate data points
for i in range(0, 25):
    plt.annotate(instagram_users[i], xy=(X_lowdim[i, 0], X_lowdim[i, 1]))

plt.show()
\end{lstlisting}

\end{document}